\documentclass[conference,a4paper]{IEEEtran}
\IEEEoverridecommandlockouts
\def\BibTeX{{\rm B\kern-.05em{\sc i\kern-.025em b}\kern-.08em
		T\kern-.1667em\lower.7ex\hbox{E}\kern-.125emX}}
	
\usepackage{amsmath,amssymb,amsfonts}
\usepackage{algorithm}
\usepackage{algpseudocode}
\usepackage{graphicx}
\usepackage{textcomp}
\usepackage{xcolor}

\usepackage[english]{babel}
\usepackage[utf8]{inputenc}
\usepackage[labelfont=bf,justification=raggedright,singlelinecheck=false,font=small]{caption}
\usepackage{placeins}
\usepackage{color}
\usepackage{mathtools}
\usepackage{xspace}
\usepackage{microtype}

\usepackage{cite} % damit Quellen bei Mehrfachzitierungen automatisch sortiert werden
\usepackage{tabularx}
\usepackage{multirow}
\usepackage{makecell}

\usepackage{gensymb} % for \degree celsius
\usepackage{booktabs}
\usepackage{ifthen}
\usepackage{eurosym} 
\usepackage{amsbsy} % bold characters in math

\usepackage[mathcal]{euscript}

\usepackage{subfigure}

\usepackage{grffile}

\usepackage{icomma}
\usepackage{tabularx}
\usepackage{nicefrac} %for smaller fraction in text mode
\usepackage{listings} % to adjust code listings
	\lstdefinestyle{myCustomMatlabStyle}{
		tabsize=4,
		showspaces=false,
		showstringspaces=false
	}
\usepackage{transparent} % was necessary because I had some transparent color in my exemplary_system_overview_small_latexFont.pdf (I think the grey, could be changed)
\usepackage{import} % to have both .pdf and .pdf_tex in a sub folder
\definecolor{myblue}{RGB}{62, 176, 247}
\usepackage[hidelinks]{hyperref} % to hide ugly red boxes

\usepackage[acronym]{glossaries} % for abbreviations and nomenclature

\makeatletter
\def\IEEElabelanchoreqn#1{\bgroup
	\def\@currentlabel{\p@equation\theequation}\relax
	\def\@currentHref{\@IEEEtheHrefequation}\label{#1}\relax
	\Hy@raisedlink{\hyper@anchorstart{\@currentHref}}\relax
	\Hy@raisedlink{\hyper@anchorend}\egroup}
\makeatother
\newcommand{\subnumberinglabel}[1]{\IEEEyesnumber
	\IEEEyessubnumber*\IEEElabelanchoreqn{#1}}

\usepackage[compact]{titlesec}
	\titlespacing{\section}{0pt}{2ex}{1ex}
	\titlespacing{\subsection}{0pt}{1ex}{0.2ex}
	
\newcommand{\T}{\ensuremath{^\intercal}\xspace}

\newcommand{\nk}[1][]{
	\ifthenelse{ \equal{#1}{} }
	{\ensuremath{(n|k)}\xspace}
	{\ensuremath{(#1|k)}\xspace}
}       
\let\k\relax % \k was already defined as some polish accent
\newcommand{\k}{\ensuremath{(k)}\xspace}
\newcommand{\kpo}{\ensuremath{(k+1)}\xspace}

\newcommand{\Np}{\ensuremath{N_\g{p}}\xspace}
\newcommand{\Npred}{\ensuremath{\Np}\xspace}

\newcommand{\sumnNp}{\sum_{n=0}^{\Np-1}}

\newcommand{\sumnpoNppo}{\sum_{n=1}^{\Np}}

\newcommand{\uvek}{\ensuremath{u}\xspace}

\newcommand{\ukk}[1][]{
	\ifthenelse{ \equal{#1}{} }
	{\ensuremath{\uvek(k+1)}\xspace}
	{\ensuremath{\uvek(k+{#1})}\xspace}
} 

\newcommand{\useq}{\ensuremath{\boldsymbol{\uvek}}\xspace}

\newcommand{\lmon}{\ensuremath{\ell_\g{mon}}\xspace}

\newcommand{\Jmon}{\ensuremath{J_\g{mon}}\xspace}

\newcommand{\Jcomf}{\ensuremath{J_\g{comf}}\xspace}

\newcommand{\Jopt}{\ensuremath{J_\g{opt}}\xspace}

\newcommand{\Jserver}{\ensuremath{J_\g{server}}\xspace}

\newcommand{\cCHP}{\ensuremath{c_\g{cur}}\xspace} % Konstante fuer CHP/ Kopplung heat und el

\newcommand{\epsc}{\ensuremath{\varepsilon_\g{c}}\xspace}

\newcommand{\Cthi}[1][]{
	\ifthenelse{ \equal{#1}{} }
	{\ensuremath{C_{\g{th},i}}\xspace}
	{\ensuremath{C_{\g{th},{#1}}}\xspace}
}        
\newcommand{\Cthj}{\ensuremath{C_{\g{th},j}}\xspace}

\newcommand{\Hai}[1][]{
	\ifthenelse{ \equal{#1}{} }
	{\ensuremath{H_{\g{air},i}}\xspace}
	{\ensuremath{H_{\g{air},{#1}}}\xspace}
} 

\newcommand{\betaij}{\ensuremath{\beta_{ij}}\xspace}

\newcommand{\cchp}{\cCHP}

\renewcommand{\sc}{\ensuremath{s_\g{c}}\xspace}

\newcommand{\Pgrid}{\ensuremath{P_\g{grid}}\xspace}

\newcommand{\Pchp}{\ensuremath{P_\g{chp}}\xspace}

\newcommand{\Pdem}{\ensuremath{P_\g{dem}}\xspace}

\newcommand{\Qrad}{\ensuremath{\dot{Q}_\g{rad}}\xspace}
\newcommand{\Qcool}{\ensuremath{\dot{Q}_\g{cool}}\xspace}
\newcommand{\Qcooli}[1][]{
	\ifthenelse{ \equal{#1}{} }
	{\ensuremath{\dot{Q}_{\g{cool,}i}}\xspace}
	{\ensuremath{\dot{Q}_\g{cool,{#1}}}\xspace}
}

\newcommand{\Qheat}{\ensuremath{\dot{Q}_\g{heat}}\xspace}
\newcommand{\Qheati}[1][]{
	\ifthenelse{ \equal{#1}{} }
	{\ensuremath{\dot{Q}_{\g{heat,}i}}\xspace}
	{\ensuremath{\dot{Q}_\g{heat,{#1}}}\xspace}
}

\newcommand{\Qotheri}[1][]{
	\ifthenelse{ \equal{#1}{} }
	{\ensuremath{\dot{Q}_{\g{other,}i}}\xspace}
	{\ensuremath{\dot{Q}_{\g{other,{#1}}}}\xspace}
}

\newcommand{\thetaa}{\ensuremath{\vartheta_\g{air}}\xspace}
\newcommand{\thetai}[1][]{
	\ifthenelse{ \equal{#1}{} }
	{\ensuremath{\vartheta_{i}}\xspace}
	{\ensuremath{\vartheta_{\g{#1}}}\xspace}
}

\newcommand{\thetabi}[1][]{
	\ifthenelse{ \equal{#1}{} }
	{\ensuremath{\vartheta_{\g{b},i}}\xspace}
	{\ensuremath{\vartheta_{\g{b,{#1}}}}\xspace}
}

\newcommand{\thetaid}{\ensuremath{\dot{\vartheta}_{i}}\xspace}

\newcommand{\thetaj}{\ensuremath{\vartheta_{j}}\xspace}

\newcommand{\Pbat}{\ensuremath{P_\g{bat}}\xspace}

\newcommand{\thetait}[1][]{
	\ifthenelse{ \equal{#1}{} }
	{\ensuremath{\vartheta_{i}(t)}\xspace}
	{\ensuremath{\vartheta_{#1}(t)}\xspace}
}

\newcommand{\thetaat}{\ensuremath{\thetaa(t)}\xspace}
\newcommand{\Qtotit}[1][]{
	\ifthenelse{ \equal{#1}{} }
	{\ensuremath{\dot{Q}_{\g{tot,}i}}\xspace}
	{\ensuremath{\dot{Q}_{\g{tot,}{#1}}}\xspace}
} 

\newcommand{\Tsamp}{\ensuremath{T_\g{s}}\xspace}

\newcommand{\scvek}{\ensuremath{E}\xspace}

\newcommand{\Qchpk}{\ensuremath{\dot{Q}_\g{chp}(k)}\xspace}

\newcommand{\Pdemk}{\ensuremath{P_\g{dem}(k)}\xspace}

\newcommand{\Qradk}{\ensuremath{\dot{Q}_\g{rad}(k)}\xspace}

\newcommand{\thetaak}{\ensuremath{\thetaa(k)}\xspace}
\newcommand{\wmon}{\ensuremath{w_\g{mon}}\xspace}

\newcommand{\wcomf}{\ensuremath{w_\g{comf}}\xspace}

\newcommand{\wserver}{\ensuremath{w_\g{server}}\xspace}

\newcommand{\wthi}[1][]{
	\ifthenelse{ \equal{#1}{} }
	{\ensuremath{w_{\g{th},i}}\xspace}
	{\ensuremath{w_{\g{th},{#1}}}\xspace}
}

\newcommand{\wthbi}[1][]{
	\ifthenelse{ \equal{#1}{} }
	{\ensuremath{w_{\g{b},i}}\xspace}
	{\ensuremath{w_{\g{b},{#1}}}\xspace}
}

\newcommand{\wthsi}[1][]{
	\ifthenelse{ \equal{#1}{} }
	{\ensuremath{w_{\g{s},i}}\xspace}
	{\ensuremath{w_{\g{s},{#1}}}\xspace}
}

\newcommand{\Imat}[2][]{
	\ifthenelse{ \equal{#1}{} }
	{\ensuremath{\boldsymbol{I}}\xspace}
	{\ensuremath{\boldsymbol{I}_{{#1}\times {#2}} }\xspace}
}
\newcommand{\Zeromat}[2][]{
	\ifthenelse{ \equal{#1}{} }
	{\ensuremath{\boldsymbol{0}}\xspace}
	{\ensuremath{\boldsymbol{0}_{{#1}\times {#2}} }\xspace}
}
\newcommand{\Onemat}[2][]{
	\ifthenelse{ \equal{#1}{} }
	{\ensuremath{\boldsymbol{1}}\xspace}
	{\ensuremath{\boldsymbol{1}_{{#1}\times {#2}} }\xspace}
}

\newcommand{\indthMono}{\g{th}} % Note: changed for retraining conference paper!
\newcommand{\xthMono}{\ensuremath{x_\indthMono}\xspace}

\newcommand{\uthMono}{\ensuremath{u_\indthMono}\xspace}
\newcommand{\dthMono}{\ensuremath{d_\indthMono}\xspace}
\newcommand{\AthMono}{\ensuremath{A_\indthMono}\xspace}
\newcommand{\BthMono}{\ensuremath{B_\indthMono}\xspace}
\newcommand{\SthMono}{\ensuremath{S_\indthMono}\xspace}

\newcommand{\indmono}{\g{all}}
\newcommand{\umono}{\ensuremath{u_\indmono}\xspace}

\newcommand{\Eevi}[1][]{
	\ifthenelse{ \equal{#1}{} }
	{\ensuremath{E_{\g{EV},i}}\xspace}
	{\ensuremath{E_{\g{EV,{#1}}}}\xspace}
}
\newcommand{\Eevarri}[1][]{
	\ifthenelse{ \equal{#1}{} }
	{\ensuremath{E_{\g{EV,arr},i}}\xspace}
	{\ensuremath{E_{\g{EV,arr,{#1}}}}\xspace}
}
\newcommand{\Eevdepi}[1][]{
	\ifthenelse{ \equal{#1}{} }
	{\ensuremath{E_{\g{EV,dep},i}}\xspace}
	{\ensuremath{E_{\g{EV,dep,{#1}}}}\xspace}
}
\newcommand{\Eevmini}[1][]{
	\ifthenelse{ \equal{#1}{} }
	{\ensuremath{E_{\g{EV,min},i}}\xspace}
	{\ensuremath{E_{\g{EV,min,{#1}}}}\xspace}
}

\newcommand{\Cevi}[1][]{
	\ifthenelse{ \equal{#1}{} }
	{\ensuremath{C_{\g{EV},i}}\xspace}
	{\ensuremath{C_{\g{EV,{#1}}}}\xspace}
}
\newcommand{\Cevarri}[1][]{
	\ifthenelse{ \equal{#1}{} }
	{\ensuremath{C_{\g{EV,arr},i}}\xspace}
	{\ensuremath{C_{\g{EV,arr,{#1}}}}\xspace}
}
\newcommand{\Cevdepi}[1][]{
	\ifthenelse{ \equal{#1}{} }
	{\ensuremath{C_{\g{EV,dep},i}}\xspace}
	{\ensuremath{C_{\g{EV,dep,{#1}}}}\xspace}
}

\newcommand{\Pevi}[1][]{
	\ifthenelse{ \equal{#1}{} }
	{\ensuremath{P_{\g{EV},i}}\xspace}
	{\ensuremath{P_{\g{EV,{#1}}}}\xspace}
}
\newcommand{\Pevmaxi}[1][]{
	\ifthenelse{ \equal{#1}{} }
	{\ensuremath{P_{\g{EV,max},i}}\xspace}
	{\ensuremath{P_{\g{EV,max,{#1}}}}\xspace}
}

\newcommand{\Devi}[1][]{
	\ifthenelse{ \equal{#1}{} }
	{\ensuremath{D_\g{i}}\xspace}
	{\ensuremath{D_{\g{#1}}}\xspace}
}

\newcommand{\mmpi}[1][]{
	\ifthenelse{ \equal{#1}{} }
	{\ensuremath{m_{\g{MP},i}}\xspace}
	{\ensuremath{m_{\g{MP},{#1}}}\xspace}
}

\newcommand{\Ppv}{\ensuremath{P_\g{PV}}\xspace}

\newcommand{\thetampi}[1][]{
	\ifthenelse{ \equal{#1}{} }
	{\ensuremath{\vartheta_{\g{MP},i}}\xspace}
	{\ensuremath{\vartheta_{\g{MP},{#1}}}\xspace}
}

\newcommand{\pmpi}[1][]{
	\ifthenelse{ \equal{#1}{} }
	{\ensuremath{p_{\g{MP},i}}\xspace}
	{\ensuremath{p_{\g{MP},{#1}}}\xspace}
}

\newcommand{\err}{\ensuremath{\epsilon}\xspace}
\newcommand{\erri}[1][]{
	\ifthenelse{ \equal{#1}{} }
	{\ensuremath{\epsilon_{i}}\xspace}
	{\ensuremath{\epsilon_{{#1}}}\xspace}
}

	\newcommand{\errk}{\ensuremath{\err\k}\xspace}
	\newcommand{\errik}{\ensuremath{\erri\k}\xspace}

\newcommand{\comp}{\ensuremath{\tilde{\epsilon}}\xspace}

	\newcommand{\compk}{\ensuremath{\comp\k}\xspace}

\newcommand{\todk}{\ensuremath{\mathrm{ToD}(k)}\xspace}
\newcommand{\dowk}{\ensuremath{\mathrm{DoW}(k)}\xspace}

\newcommand{\WMARE}{\ensuremath{\mathrm{WMARE}}\xspace}

\newcommand{\Pserver}[1]{\ensuremath{P_{\g{server},#1}}\xspace}

\let\st\relax
\DeclareMathOperator*{\st}{s.t.}

\newcommand{\eg}{e.\thinspace{}g.\@\xspace}
\newcommand{\ie}{i.\thinspace{}e.\@\xspace}

\newcommand{\pagerefh}[1]{\hyperref[#1]{page~\pageref*{#1}}}
\newcommand{\secref}[1]{\hyperref[#1]{Section~\ref*{#1}}}
\newcommand{\chapref}[1]{\hyperref[#1]{Chapter~\ref*{#1}}}
\newcommand{\figref}[1]{Figure~\ref{#1}}
\newcommand{\tabref}[1]{\hyperref[#1]{Table~\ref*{#1}}}

\newcommand{\g}{\mathrm}

\newcommand{\kW}{\ensuremath{\g{kW}}\xspace}
\newcommand{\kWh}{\ensuremath{\g{kWh}}\xspace}

\newcommand{\kWpK}{\ensuremath{\frac{\g{kW}}{\g{K}}}\xspace}

\newcommand{\kWhpK}{\ensuremath{\frac{\g{kWh}}{\g{K}}}\xspace}

\newcommand{\kWel}{\ensuremath{\g{kW}_\g{el}}\xspace}

\newcommand{\kWpeak}{\ensuremath{\g{kWp}}\xspace}

\newcommand{\degC}{\ensuremath{\degree \g{C}\xspace}}
\newcommand{\kelvin}{\ensuremath{\g{K}\xspace}}

\newcommand{\qm}{\ensuremath{\g{m}^2}\xspace}

\newcommand{\hours}  {\ensuremath{\,\g{h}}\xspace}

\newcommand\MEUR[1]{\if@EURleft\text{\euro}\,\fi#1\if@EURleft\else\,\text{\euro}\fi}

\newcommand{\percent}{\,\%\xspace}

\makeglossaries

\newacronym{bem}{BEM}{building energy management}
\newacronym{res}{RES}{renewable energy source}
\newacronym{mpc}{MPC}{Model Predictive Control}
\newacronym{moo}{MOO}{multi-objective optimization}
\newacronym{ev}{EV}{electric vehicle}
\newacronym{bps}{BPS}{building performance software}
\newacronym{ode}{ODE}{ordinary differential (or difference) equation}
\newacronym{hri}{HRI}{Honda Research Institute}
\newacronym{hvac}{HVAC}{heating, ventilation, and air conditioning}
\newacronym{chp}{CHP}{combined heat and power plant}
\newacronym{pv}{PV}{photovoltaic}
\newacronym{empc}{EMPC}{Economic Model Predictive Control}
\newacronym{der}{DER}{distributed energy resource}
\newacronym{ess}{ESS}{energy storage system}
\newacronym{tes}{TES}{thermal energy storage}
\newacronym{ocp}{OCP}{optimal control problem}
\newacronym{lp}{LP}{linear programming}
\newacronym{qp}{QP}{quadratic programming}
\newacronym{nlp}{NLP}{nonlinear programming}
\newacronym{milp}{MILP}{mixed-integer linear programming}
\newacronym{minlp}{MINLP}{mixed-integer nonlinear programming}
\newacronym{dm}{DM}{decision maker}
\newacronym{awds}{AWDS}{adaptive weight determination scheme}
\newacronym{nbi}{NBI}{normal boundary intersection}
\newacronym{chim}{CHIM}{convex hull of individual minima}
\newacronym{fpbi}{FPBI}{focus point boundary intersection}
\newacronym{cup}{CUP}{closest to Utopia point}
\newacronym{aep}{AEP}{angle to the extreme points}
\newacronym{atn}{ATN}{angle to the neighbor points}
\newacronym{ci}{CI}{carbon intensity}
\newacronym{hl}{HL}{higher level}
\newacronym{ll}{LL}{lower level}

\newacronym[\glslongpluralkey={Gaussian Processes}]{gp}{GP}{Gaussian Process}
\newacronym{bepst}{BEPST}{building energy performance simulation tool}
\newacronym{ann}{ANN}{artificial neural network}
\newacronym{rnn}{RNN}{recurrent neural network}
\newacronym{ems}{EMS}{energy management system}
\newacronym{svr}{SVR}{support vector regression}
\newacronym{dr}{DR}{demand response}
\newacronym{ga}{GA}{Genetic Algorithm}
\newacronym{rc}{RC}{Resistor-Capacitor}
\newacronym{rbf}{RBF}{radial basis function}
\newacronym{pmv}{PMV}{predicted mean vote}
\newacronym{ekf}{EKF}{extended Kalman filter}
\newacronym{ml}{ML}{machine learning}
\newacronym{lstm}{LSTM}{long short-term memory}
\newacronym{pcnn}{PCNN}{physically consistent neural network}
\newacronym{pinn}{PINN}{physics-informed neural network}
\newacronym{knn}{k-NN}{k-nearest neighbors}
\newacronym{mlp}{MLP}{multilayer perceptron}

\newacronym{epl}{EPL}{expected performance loss}
\newacronym{sil}{SiL}{software-in-the-loop}
\newacronym{mae}{MAE}{mean absolute error}
\newacronym{wmare}{WMARE}{weighted mean absolute residual error}
\newacronym{wmre}{WMRE}{weighted mean residual error}
\newacronym{ware}{WARE}{weighted absolute residual error}
\newacronym{rmse}{RMSE}{root mean square error}
\newacronym{fmu}{FMU}{functional mock-up unit}

\glsdisablehyper

\begin{document}

\title{
	Evaluating the Impact of Data Availability on Machine Learning-augmented MPC for a Building Energy Management System
}

\author{
	Jens Engel$^{1,2}$,
	Thomas Schmitt$^{1}$,
	Tobias Rodemann$^{1}$,
	Jürgen Adamy$^{2}$
	\thanks{$^{1}$Honda Research Institute Europe GmbH, 
		Offenbach, Germany. 
		E-mail: {\tt\footnotesize \{jens.engel, thomas.schmitt, tobias.rodemann\}@honda-ri.de}
	}%
	\thanks{$^{2}$Control Methods and Intelligent Systems Laboratory, 
		Technical University of Darmstadt, Darmstadt, Germany
		E-mail: {\tt\footnotesize juergen.adamy@tu-darmstadt.de}
	}%
}

\maketitle

\begin{abstract}
	A major challenge in the development of \gls{mpc}-based \glspl{ems} for buildings is the availability of an accurate model.
	One approach to address this is to augment an existing gray-box model with data-driven residual estimators.
	The efficacy of such estimators, and hence the performance of the \gls{ems}, relies on the availability of sufficient and suitable training data.
	In this work, we evaluate how different data availability scenarios affect estimator and controller performance.
	To do this, we perform \gls{sil} simulation with a physics-based digital twin using real measurement data.
	Simulation results show that acceptable estimation and control performance can already be achieved with limited available data, and we confirm that leveraging historical data for pretraining boosts efficacy.
\end{abstract}

\begin{IEEEkeywords}
	data-driven residual estimator, model predictive control, digital twin, building control, machine learning
\end{IEEEkeywords}

% reset glossary to reintroduce abbreviations from abstract
\glsresetall
%
%%%%%%%%%%%%%%%%%%%%%%%%%%%%%%%%%%%%%%%%%%%%%%%%%%%%%%%%%%%%%%%%%%%%%%%%%%%%%%%%

\section{Introduction}
The increasing influx of \glsdesc{res} into the public power grid leads to an increasing demand for intelligent \glspl{ems} for buildings.
A promising control method for such \glspl{ems} is \gls{mpc} \cite{drgona2020all, pean2019price}.
Yet, for \gls{mpc}-based \gls{ems} control approaches to be effective, an accurate underlying model is needed.

Various approaches for model building exist, generally categorized into white-box, gray-box and black-box modeling.
White-box models build using \eg \acrlongpl{bepst} like EnergyPlus or TRNSYS can provide very high levels of detail and accuracy, but can generally not be directly used within an \gls{mpc}'s \gls{ocp}.
Grey-box models, like state space or \gls{rc} models, have a lower level of accuracy, but are well suited for the use within an \gls{ocp}.
Both white-box and gray-box modeling of buildings are highly complex and require expert knowledge specific to each building, \ie models cannot be easily transferred to other buildings \cite{drgona2020all, privara2013building, tian2018review}.
Hence, data-driven black box modeling has become a popular alternative in recent years \cite{wang2019data, maddalena2020data, halhoulmerabet2021intelligent}, e.g., using \glspl{gp}, \glspl{ann},  reinforcement learning, or, notably, using \glspl{pinn} \cite{gokhale2022physics, dinatale2022physically, wang2023physics}.
While such black-box models require only a relatively low modeling effort, a substantial amount of data is needed for training, which can be challenging to acquire \cite{maddalena2020data, halhoulmerabet2021intelligent}.
At the same time, the ability to incorporate specific domain knowledge, interpretability, and adaptability of the model, which are the strengths of white/gray-box models, are lost.
Furthermore, simple linear gray-box models have been shown to achieve similar performance \cite{buenning2022physics, stoffel2024real}.
Hence, hybrid approaches combining these paradigms could be a promising way to enable larger scale real world adoption of building \glspl{ems} \cite{halhoulmerabet2021intelligent}.

One such hybrid approach is the augmentation of a gray-box model using a  \textit{residual estimator}.
Instead of replacing or adapting the gray-box model itself, its model error is reduced by estimating the residual value using an additional data-driven model.
This concept has been applied to various applications of \gls{mpc} \cite{picotti2024real, scheurenberg2023evaluation, massagray2018hybrid}.
In the context of building \glspl{ems}, a common approach is to use residual estimators to directly predict heat disturbances acting on the building, caused through occupant behavior and ambient effects
\cite{
	thilker2021advanced,
	ellis2021machine,
	kumar2023grey,
	schmitt2023regression,
	liang2024energy,
	engel2024hierarchical
}.
These exogenous effects represent one of the largest error sources in buildings \cite{drgona2020all,oldewurtel2013importance}, and these approaches have shown to be able to significantly reduce the residual error \cite{schmitt2023regression}.
%One advantage of the estimation of only exogenous effects is that integration into the \gls{ocp} is straight-forward, as the estimator does not rely on any decision variables.

In previous work, we have shown that the control performance of an \gls{mpc} augmented by machine learning-based residual estimation relies strongly on the availability and performance of the estimators \cite{engel2024hierarchical}.
As such estimators rely on training data, it becomes essential to investigate how the availability of data correlates with estimator performance.
When applying such an approach in a real-world setting, historical data for pre-training estimators may not be available.
This motivates two questions:
1.) How much data is needed until estimators achieve a good performance?
2.) If historical data is available and new data is collected in-the-loop, how should this new data be incorporated and the estimators retrained?
While many authors note that these are important issues to be addressed \cite{liang2024energy, ellis2021machine, schmitt2023regression}, to the best of our knowledge, none have evaluated this systematically.
In this study, we want to evaluate these questions in the context of a \gls{mpc}-based \gls{ems} of an industrial multi-zone building which integrates error compensation through data-driven heat disturbance estimation.
In \gls{sil} simulation with a physics-based digital twin, we evaluate different data availability scenarios and retraining strategies.
We analyze both the accuracy of the residual estimator in the different settings, as well as how this relates to the performance of the proposed control approach.

This paper is organized as follows:
The building under study, its digital twin, as well as the proposed control approach are described in \secref{sec:ems}.
The residual error estimation is described in \secref{sec:error_compensation}.
The simulation study and the discussion of the results is presented in \secref{sec:simulation_study}.
Finally, in \secref{sec:conclusion}, a conclusion is given.
\section{Energy Management System}
\label{sec:ems}
In this work we consider a medium-sized company building situated in Offenbach, Germany.
It has a total footprint of approx.\ $13,\,000\,\qm$, which is partitioned into 9 temperature zones,
namely offices, halls, workshops, an emissions lab and server rooms. 
In addition to the public power grid connection, it has a gas-fired \gls{chp} for co-production of heat and electricity with $199\,\kWel$, 
and a \gls{pv} plant of $750\,\kWpeak$, which supply an average electrical load demand of approx. $250\,\kW$.
Besides the heat produced by the \gls{chp}, thermal energy is supplied by gas-fired heating boilers and an electric \gls{hvac} system.
Furthermore, as electrical energy storage, a second-life battery with a capacity of $98\,\kWh$ is available \cite{stadie2021v2b}.
%Furthermore, a second-life battery with a capacity of $98\,\kWh$ is available \cite{stadie2021v2b}.

As a surrogate for the real building, we use a digital twin of the facility.
The digital twin is a physics-based white-box model implemented in SimulationX \cite{unger2012green, castellani2021real}.
It covers the temperature zones' connections and heat losses to both the ambient air and the ground.
It takes into account internal heat gains from electrical usage and building occupancy,
and features a detailed \gls{hvac} system.
It makes use of real measurement data of electrical loads and weather.
For the control of the building, an \gls{mpc} control approach adopted from \cite{engel2024hierarchical} is used.

\subsection{Modeling}
\label{sec:ems_modeling}
For this study, we use a time-discrete state-space model. 
While a hierarchical approach can be used to improve scalability of large buildings \cite{engel2022hierarchical}, we choose a monolithic model for simplicity. 
Note that the results presented in \secref{sec:simulation_study} are very similar for the hierarchical approach. 
As states, the model considers 
the stationary battery's stored energy $\scvek$ in $\kWh$,  
and the 9 zone temperatures $\thetai$ in $\degC$, 
where $\thetai[1]$ -- $\thetai[7]$ are regular building zones with heating and cooling, 
and $\thetai[8]$ and $\thetai[9]$ represent server rooms, which have no heating systems. 

As inputs to the system, we consider
the grid power \Pgrid, 
the (electrical) \gls{chp} power \Pchp, 
the battery's (dis-)charging power \Pbat, 
the gas heating power \Qrad, 
and the \gls{hvac} cooling power \Qcool.
The heating and cooling powers are split into 
$\Qheati~\forall\, i = 1,\,\ldots\,, 7$ 
and $\Qcooli~\forall\, i = 1,\,\ldots\,, 9$. 
As disturbances, we consider 
the \gls{pv} power \Ppv, 
the building's electrical load demand \Pdem, 
the ambient air temperature \thetaa, 
and (constant) heat disturbances $\Qotheri$, 
which are losses to the ground for zones for $i = 1,\,\ldots\,,7$ % there should be a space between the "," and the "7", but then we have 1 more line 
and internal heatings for $i = 8, 9$.
The unit of all electrical and thermal powers is \kW. 

The \textit{time-continuous} differential equation of a temperature zone is thus given by
%\begin{IEEEeqnarray*}{rCl}
%	\thetaid(t) &=& 
%	\frac{1}{\Cthi} \left( \Qheati(t) + \Qcooli(t) + \Qotheri(t) \right) \IEEEnonumber    
%	%	\IEEEeqnarraynumspace
%	\\
%	&& 
%	-\> \sum_{j \neq i} \frac{\betaij}{\Cthi} \left( \thetai(t) - \thetaj(t) \right) \IEEEnonumber \\
%	%	\IEEEeqnarraynumspace
%	&& 
%	- \frac{\Hai}{\Cthi} \left( \thetai(t) - \thetaat \right), \label{eq:ode_theta_i} \IEEEyesnumber
%	%	+ \erri(t)
%	%	\IEEEeqnarraynumspace
%\end{IEEEeqnarray*} 
%\begin{IEEEeqnarray*}{rcl}
%	\thetaid(t) &=& 
%	\frac{1}{\Cthi} \left( \Qheati(t) + \Qcooli(t) + \Qotheri(t) \right) \IEEEyesnumber    
%	%	\IEEEeqnarraynumspace
%	\\
%	&& 
%	-\> \sum_{j \neq i} \frac{\betaij}{\Cthi} \left( \thetai(t) - \thetaj(t) \right) 
%%	\IEEEnonumber \\
%	%	\IEEEeqnarraynumspace
%%	&& 
%	- \frac{\Hai}{\Cthi} \left( \thetai(t) - \thetaat \right), \label{eq:ode_theta_i} \IEEEnonumber
%	%	+ \erri(t)
%	%	\IEEEeqnarraynumspace
%\end{IEEEeqnarray*} 
\begin{IEEEeqnarray*}{rl}
	\thetaid(t) &= 
	\frac{1}{\Cthi} \left( \Qheati(t) + \Qcooli(t) + \Qotheri(t) \right) \IEEEyesnumber    
	%	\IEEEeqnarraynumspace
	\\
	& 
	-\> \sum_{j \neq i} \frac{\betaij}{\Cthi} \left( \thetai(t) - \thetaj(t) \right) 
%	\IEEEnonumber \\
	%	\IEEEeqnarraynumspace
%	&& 
	- \frac{\Hai}{\Cthi} \left( \thetai(t) - \thetaat \right), \label{eq:ode_theta_i} \IEEEnonumber
	%	+ \erri(t)
	%	\IEEEeqnarraynumspace
\end{IEEEeqnarray*} 
%%%%
%\begin{IEEEeqnarray*}{rCl}
%	\thetaid(t) &=& 
%	\frac{1}{\Cthi} \left( \Qheati(t) + \Qcooli(t) + \Qotheri(t) \right) \IEEEnonumber \\
%	&& 
%	-\> \sum_{j \neq i}\! \frac{\betaij}{\Cthi}\! \left(\! \thetai(t) \!-\! \thetaj(t)\! \right) 
%	\!-\! \frac{\Hai}{\Cthi}\! \left(\! \thetai(t)\! -\! \thetaat \!\right), \label{eq:ode_theta_i} \IEEEyesnumber
%\end{IEEEeqnarray*} 
$\Cthi$ is the thermal capacity of zone $i$ in $\kWhpK$, 
$\betaij$ the heat transfer coefficient between zones $i$ and $j$ in $\kWpK$, 
and $\Hai$ the heat transfer coefficient between zone $i$ and the ambient air in $\kWpK$.
Note that $\Qheati(t) = 0~\forall\,i = 8, 9$. 
%The numerical values of all building parameters can be found in \tabref{tab:thermal_parameters}.
The numerical values of all building parameters can be found in \cite{schmitt2023regression}.
With \eqref{eq:ode_theta_i}, the thermal model part can be summarized as a (time-continuous) state space model as described in \cite[pp.\,22]{schmitt2022multi}, but omitted here for brevity. 
With the state vector 
$\xthMono = \begin{bmatrix} \thetai[1] & \cdots & \thetai[9] \end{bmatrix}\T$, 
the input vector 
$\uthMono = \begin{bmatrix} 
	\Qheati[1] & \!\!\cdots\!\! & \Qheati[7] 
	& \Qcooli[1] & \!\!\cdots\!\! & \Qcooli[9] 
	\end{bmatrix}\T$, 
and the disturbance vector 
$\dthMono = \begin{bmatrix} 
	\thetaa & \Qotheri[1] & \cdots & \Qotheri[9]
	\end{bmatrix}\T$, 
it can be discretized with the sampling time $\Tsamp = 0.5\,\g{h}$ to 
\begin{IEEEeqnarray}{rCl}\label{eq:HLab_dis_ss}
	\xthMono(k+1) &=& \AthMono(\Tsamp)\xthMono(k) + \BthMono(\Tsamp)\uthMono(k) \IEEEnonumber \\ 
	&& +\> \SthMono(\Tsamp)\dthMono(k) + \err\k. \IEEEeqnarraynumspace
	\label{eq:HLabc_agg_ss}
\end{IEEEeqnarray}
As initially motivated, we consider a residual error of the state-space model 
$\errk = \begin{bmatrix} \erri[1]\k & \cdots & \erri[9]\k \end{bmatrix}\T$ 
as a time-variant heat disturbance, 
representing exogenous influences caused by occupant behavior and ambient effects. 
The estimation of \errk is discussed in \secref{sec:error_compensation}.
The electrical part of the building can directly be expressed in time-discrete form as 
\begin{IEEEeqnarray}{rCl}
	\scvek\kpo & = & \scvek\k + \Pbat\k \cdot \Tsamp, \label{eq:scvek_ode} 
\end{IEEEeqnarray}
with the balance equation 
\begin{IEEEeqnarray}{rCl}
	0 & = & \Pgrid\k - \Pbat\k + \Pchp\k + \frac{1}{\epsc}\Qcool\k \IEEEnonumber \\ 
		& &	+\> \Pdem\k + \Ppv\k, \label{eq:balance_equation}
	\label{eq:balance_eq} 
\end{IEEEeqnarray}
where $\epsc = 1.78$ is the energy efficiency ratio of the \gls{hvac}'s cooling system. 
Thermal and electrical powers are coupled by 
\begin{IEEEeqnarray}{rCl}
	\Pchp\k & = & \cchp \cdot \Qchpk, 			\label{eq:couplings_1}\\ 
	\Qheat\k & = & \Qradk + \Qchpk, 			\label{eq:couplings_2}\\ 
	\Qheat\k & = & \sum\nolimits_{i=1}^{i=7} \Qheati\k,  \label{eq:couplings_3}\\ 
	\Qcool\k & = & \sum\nolimits_{i=1}^{i=9} \Qcooli\k,  \label{eq:couplings_4} 
\end{IEEEeqnarray}
where $\cchp = 0.677$ is the \gls{chp}'s power-to-heat ratio.

\subsection{Control Approach}
The model described in \secref{sec:ems_modeling} is utilized to control the digital twin within a \gls{sil} setting with \gls{mpc}. 
The optimization objective includes three cost functions,
{\allowdisplaybreaks
\begin{IEEEeqnarray}{rcl}
	\Jcomf(k) & = & \sumnpoNppo \sum_{i=1}^{7} \wthbi \cdot \left( \thetai\nk - 22\degC \right)^2, \label{eq:lcomf} \IEEEeqnarraynumspace \\
	\Jserver(k) & = & \sumnpoNppo  \sum_{i=8}^{9} 
	%								\bigg( 
			\wthsi  
			\Big( 
			\max\left( 15\degC - \thetai\nk,~0 \right) 
			\IEEEnonumber \IEEEeqnarraynumspace 
			\\
			&& \hphantom{\sumnpoNppo} \hphantom{\sum_{i=8}^{9}} +\> \max\left( \thetai\nk - 21\degC,~0  \right)  
			\Big),  
			%								\bigg)  
	\IEEEeqnarraynumspace \label{eq:Jserverdis} \\
%	&& \hphantom{\sumnpoNppo} +\> \max\left( \thetaserver\nk - 21\degC,~0  \right) \IEEEeqnarraynumspace \label{eq:Jserveragg}  \\ 
	\Jmon(k) & = &\!\sumnNp\!\lmon\!\left(\!\Pgrid\nk,\Pchp\nk,\Qheat\nk\!\right), \IEEEeqnarraynumspace \label{eq:Jmon} 
\end{IEEEeqnarray}
where $\wthbi$ and $\wthsi$ are weights derived from the zones' thermal capacities, \ie
\begin{IEEEeqnarray}{rCl}
	\wthbi & = & \frac{ \Cthi}{\sum_{j=1}^{7} \Cthj} ~\forall\,i=1,\,\ldots\,, 7, \label{eq:wthbi} \\
	\wthsi & = & \frac{ \Cthi}{\sum_{j=8}^{9} \Cthj} ~\forall\,i=8,\ 9,  \label{eq:wthsi}
\end{IEEEeqnarray}
and \lmon describes the costs arising from gas usage and buying/selling electrical energy from/to the public grid, including peak costs.
%\\ \todo{should we describe the intention of \Jserver and/or \Jcomfagg? $\rightarrow$ Maybe if there is space left.}
The notation $a\nk$ refers to the value of $a(k+n)$ as predicted at time step $k$, and will be used from here on.
The overall objective function is then given by 
\begin{IEEEeqnarray}{rCl}
	\label{eq:J_opt}
	\Jopt\k & = & \wcomf \cdot \Jcomf\k + \wserver \cdot \Jserver\k \IEEEnonumber \\ 
			&	& +\> \wmon \cdot \Jmon\k, 
\end{IEEEeqnarray}
where $\wcomf = 0.99$, ${\wserver = 1}$, $\wmon = 0.01$ are chosen such that the emphasis is completely on the thermal regulation, which makes the interpretation of the simulation results easier. 
Note that multi-objective \gls{mpc} can be used to either determine appropriate fixed weights \cite{schmitt2020application,schmitt2020multi}, or to dynamically choose the Pareto optimal solution of your preference \cite{schmitt2022incorporating}. 

Summarizing all decision variables as $\umono = [\uthMono\T, \Pgrid, \Pchp, \Pbat, \Qcool, \Qheat]$, 
the \gls{mpc}'s \gls{ocp} is given by
\begin{IEEEeqnarray}{rCl}
	\subnumberinglabel{eq:opt_prob_Aggregator}
	\min_{\useq_\g{\indmono}\k} & ~ & \Jopt\k 
	%	\IEEEnonumber
	\IEEEeqnarraynumspace  
	\\
	\st & & 
	\eqref{eq:HLab_dis_ss} \text{--} \eqref{eq:couplings_4},
%			\eqref{eq:HLab_dis_ss},
%			\eqref{eq:scvek_ode},
%			\eqref{eq:balance_equation},
%			\eqref{eq:couplings_1},
%			\eqref{eq:couplings_2},
%			\eqref{eq:couplings_3},
%			\eqref{eq:couplings_4},
	 ~\forall\, n = 0,\,\ldots\,, \Npred -1,   
%	 \\
%	&& \eqref{eq:constraints_HL_states_all} ~\forall\, n = 1 \ldots \Npred,   
\end{IEEEeqnarray} 
and subject to box constraints on both states and control inputs. 
$\useq_\g{\indmono}\k = \left( \umono(0|k),\, \ldots\,,\,  \umono(\Npred-1|k)\right)$
denotes the sequence of control inputs.
The time step notation $(k)$ and $(k+1)$ in \eqref{eq:HLab_dis_ss} -- \eqref{eq:couplings_4} are to be read as $(n|k)$ and $(n+1|k)$, respectively.
As the prediction horizon, we use $\Npred = 48$ steps of $\Tsamp=0.5\hours$ each, \ie 1 day in total. 

\section{Residual Error Estimation}
\label{sec:error_compensation}
As motivated in the introduction, and described in the previous section,
we consider a residual model error $\errk$ in the controller.
We assume the time-variant error to represent unknown exogenous influences.
While these influences are not directly measurable and difficult to model explicitly, they are strongly correlated with measurable and predictable data.
Therefore, we use machine learning regression models to learn a data-driven estimator $\compk$.
As features, we propose the total building load $\Pdemk$,
the time of day $\todk: \mathbb{N} \rightarrow [0, 24)$ in $\hours$,
the day of week $\dowk: \mathbb{N} \rightarrow \{0, \dots, 6\}$ in $\g{d}$,
(lags of) the ambient temperature $\thetaa(k-n)$,
as well as the electrical loads in the server zones $\Pserver1\k, \Pserver2\k$, which are part of the overall building load $\Pdem\k$ and can be metered separately.
We split the estimation into two parts: one regressor estimates the errors of the building zones $\erri[1]\k, \dots, \erri[7]\k$, a second those of the server zones $\erri[8]\k, \erri[9]\k$.
As input features for the regressors we use:
\begin{itemize}
	\item Building zones: $\Pdemk$, $\thetaak$, $\thetaa(k-1)$, $\thetaa(k-2)$, $\todk$, $\dowk$,
	\item Server zones: $\Pdemk$, $\thetaak$, $\Pserver1\k$, $\Pserver2\k$.
\end{itemize}
The target variable of the regression models is the observed model error $\err\k$.
It can be calculated as the difference between measured and predicted state.
In general, the predicted state $\xthMono\nk[1]$ may already contain an estimate  $\compk$, so the true target variable is calculated as
\begin{IEEEeqnarray}{rCl}
	\err\k & = & \big(\xthMono\kpo - \xthMono\nk[1]\big) + \compk.
\end{IEEEeqnarray}
In general, this difference will also include prediction errors of the disturbances.
As we only want to estimate exogenous influences, this can be corrected by first recalculating $\xthMono\nk[1]$ with measurements of the disturbances.
\section{Simulation Study}
\label{sec:simulation_study}
We want to evaluate how different data availability scenarios affect the performance of the previously introduced residual estimation,
and how this relates to controller performance.
To do this, we carry out a \gls{sil} simulation study using the digital twin model and real measurement data for the years 2022 and 2023.
All simulations were implemented in Python, with the digital twin connected through a \gls{fmu}, and using Gurobi as a solver.
In a first step, we simulated the controller without any error compensation, \ie $\compk = 0$, for the year 2022.
We then use this data to evaluate 4 different training data availability scenarios in simulation of the year 2023:
\begin{itemize}
	\item S0: No error compensation is applied, \ie $\compk = 0$.
	\item S1: Monthly retraining without any prior data, i.e. we incrementally simulate each month, after which we train estimators on the collected data, and apply them in the following month.
	\item S2: Analogous monthly retraining but with prior data, i.e. additionally historical training data for all of 2022 is appended to the collected data.
	\item S3: Analogous to S2, but a rolling 12 calendar month window is applied to the training data.
\end{itemize}
We compare the results against two baseline scenarios:
\begin{itemize}
	\item S4: Estimators trained on  historical full year 2022 data are applied to 2023 simulation without any retraining.
	\item S5: Estimators trained on S0, re-simulating 2023 with active compensation.
\end{itemize}
We performed the evaluation with different candidate regression models, namely XGBoost \cite{chen2016xgboost}, a \gls{mlp}, and multiple linear regression.
Out of these, XGBoost performed best for estimating the building zone errors, while the \gls{mlp} yielded very similar performance.
For the server zone estimator, a linear regressor was chosen.
Since we want to focus on assessing the general estimation performance of the intrinsic model error,
we use perfect predictions for PV power, the building’s loads and ambient air temperature.
\begin{figure}
	\captionsetup{belowskip=-10pt}
	\centering
	\includegraphics[width=\columnwidth]{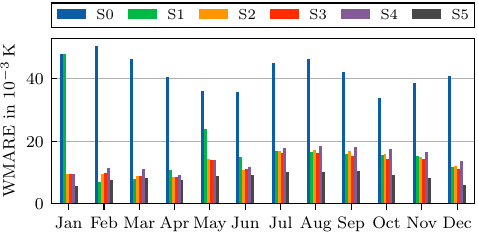}
	\caption{
		\Acrfull{wmare} over the course of the year 2023 in the different scenarios.
	}
	\label{fig:weighted_mae_over_time}
\end{figure}
\begin{figure}
	\captionsetup{belowskip=-15pt}
	\centering
	\includegraphics[width=\columnwidth]{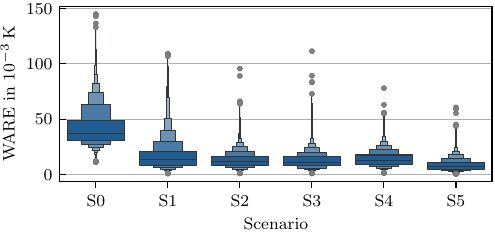}
	\caption{
		Letter-value plot showing the distribution of the \acrfull{ware} compared between the different scenarios.
%		The letter-value plot displays a series of quantiles for each dataset. 
		The central quantile represents $50\percent$ of the data, which are then halved at every next quantile, \ie $25\percent$, $12.5\percent$, and so forth.
	}
	\label{fig:residual_distribution_over_time}
\end{figure}
\figref{fig:weighted_mae_over_time} shows the \gls{wmare}, \ie
\begin{IEEEeqnarray}{rCl}
	\WMARE & = & \frac{1}{N} \sum_{k=0}^{N-1} {\underbrace{\sum_{i=1}^{9} \wthi \cdot |\errik|}_{\g{WARE}\k}}, \IEEEeqnarraynumspace
\end{IEEEeqnarray}
%\begin{IEEEeqnarray}{rCl}
%	\WMARE & = & \sum_{k=0}^{N-1} \frac{1}{N} \sum_{i=1}^{9} \wthi \cdot |\errik|,
%\end{IEEEeqnarray}
with $\wthi = \frac{ \Cthi}{\sum_{j=1}^{9} \Cthj} ~\forall\,i=1, \ldots, 9$
over the course of the year for all scenarios.
It shows that even without any prior training data in scenario S1, good estimation performance can be achieved in the first months of the year,
even very slightly outperforming the pretraining scenarios S2--S5.
During May, the estimation performance drops significantly, recovering in June.
From July on, S1 achieves similar performance to S2 and S3, indicating that half a year of training data is a critical amount for achieving consistent performance.
Scenario S3 seems to slightly outperform S2, suggesting that forgetting data from the previous year may be advantageous.
Scenario S4, \ie when no retraining of the previous year estimator is applied, exhibits similar but decreased performance to scenarios S2 and S3.
Overall, scenarios S1--S4 have good but, compared to the perfect information scenario S5, significantly decreased performance.
These findings are confirmed by the distribution of the \gls{ware} over the whole year 2023 as shown in \figref{fig:residual_distribution_over_time}.
%It shows a letter-value plot, which displays a series of quantiles for a dataset. 
%Starting with the central $50\percent$ of data, it halves the quantile interval at each step, \ie $25\percent$, $12.5\percent$, and so forth.
Without available prior training data, the worst compensation performance is achieved, but the overall error is still significantly reduced (S1).
The errors of S2 and S3 are almost equally distributed, showing no significant difference.
Even though the performance of S4 is worse than S2 and S3, the overall error is still significantly reduced, indicating that the applied regressor models generalize reasonably well.
A key criterion of the thermal control performance is the \gls{rmse} of the temperature tracking,
which indicates how far individual zones deviate from the reference temperature,
\begin{IEEEeqnarray}{rCl}
	\g{RMSE} = \sum_{i=1}^7 \wthbi \cdot \sqrt{\frac{1}{N} \sum_{k=1}^{N} \left(\thetai\k - 22\degC \right)^2}.
\end{IEEEeqnarray}
\begin{figure}
	\captionsetup{belowskip=-7pt}
	\centering
	\includegraphics[width=\columnwidth]{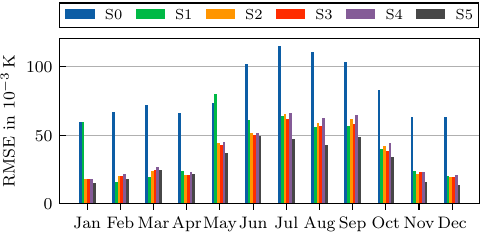}
	\caption{
		\Acrfull{rmse} of the temperature tracking over the course of the year 2023 in the different scenarios.
	}
	\label{fig:tracking_error_over_time}
\end{figure}
\figref{fig:tracking_error_over_time} shows this metric over the course of the year 2023.
It can be seen that the difference in tracking performance between scenarios mostly matches that in compensation performance.
While the latter is relatively consistent from June onward,
the tracking performance has a seasonal trend.
This is due to the controller weighing the tracking error against monetary costs (especially peak costs).
This is also the reason for why there is almost no difference in tracking performance in June: under certain conditions, to avoid peaks, better compensation performance may not significantly improve the tracking error.
Notably, in May, in S1 a worse tracking performance than in S0 is achieved, even though the absolute residual error was significantly reduced.
This is due to the fact that while the mean \textit{absolute} error was reduced, the mean error has a significant bias, as shown in \figref{fig:weighted_mean_error_over_time}.
Without historical data, the estimator cannot account for the vanishing negative bias in May.
This causes the temperature to drift, as the controller will consistently receive biased estimates of the heat disturbances.
\begin{figure}
	\captionsetup{belowskip=-15pt}
	\centering
	\includegraphics[width=\columnwidth]{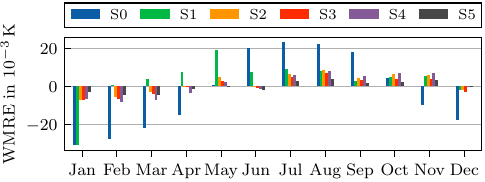}
	\caption{
		\Acrfull{wmre} over the course of the year 2023 in the different scenarios.
	}
	\label{fig:weighted_mean_error_over_time}
\end{figure}

To confirm the overall findings, we performed the same simulation study for the year 2022 (by simulating 2021 first).
\tabref{tab:overall_results} shows the full-year \gls{wmare} and the \gls{rmse} of the temperature tracking for both 2022 and 2023.
These data confirm the previous findings that incremental learning without prior data can quickly achieve good performance and that leveraging historical data is beneficial.
Furthermore, the data indicate that there is a slight advantage if windowed historical data is used for training.
\begin{table}
	\caption{\Acrfull{wmare} and \acrfull{rmse} of the tracking performance for all scenarios in 2022 and 2023.}
	\begin{tabularx}{\columnwidth}{llrrrrrr}
		\toprule
		&  & S0 & S1 & S2 & S3 & S4 & S5 \\
		Metric & Year & &   &  &  &  &  \\
		\midrule
		\gls{wmare} & 2022 & 41.21 & 12.96 & 11.42 & 11.38 & 12.63 & 7.58 \\
		in $10^{-3}\,\kelvin$ & 2023 & 40.38 & 14.98 & 13.69 & 13.09 & 14.89 & 8.85 \\
		\midrule
		\gls{rmse} & 2022 & 97.35 & 53.79 & 48.92 & 48.68 & 51.16 & 44.62 \\
		in $10^{-3}\,\kelvin$ & 2023 & 89.47 & 51.07 & 46.40 & 44.56 & 48.15 & 37.56 \\
		\bottomrule
	\end{tabularx}
	\vspace{-2em} 
	\label{tab:overall_results}
\end{table}

\section{Conclusion \& Outlook}
\label{sec:conclusion}
In this study we evaluated how different data availability scenarios affect the performance of residual estimation of exogenous influences on an industrial multi-zone building and how this relates to controller performance.
We found that even without historical data available, data-driven residual estimation can quickly achieve good performance on a very conservative amount of data.
If historical data is available, building on this data in an incremental fashion improves estimation performance.
We further showed that thermal control performance of our proposed approach benefits from estimation accuracy, where higher accuracy leads to better performance.
Overall, we find that the proposed residual error estimation approach can achieve good results on a conservative amount of training.

%%%%%%%%%%%%%%%%%%%%%%%%%%%%%%%%%%%%%%%%%%%%%%%%%%%%%%%%%%%%%%%%%%%%%%%%%%%%%%%%

\FloatBarrier
% If space is sparse: 
%\newcommand{\BIBdecl}{\setlength{\itemsep}{-6pt}}
\bibliographystyle{IEEEtran}
\bibliography{common/lib_error_comp}
	
\end{document}